# Solar energy harvesting in magnetoelectric coupled manganese ferrite nanoparticles incorporated nanocomposite polymer films

Sonali Pradhan[1,3], Pratik P. Deshmukh[1,3], S. N. Jha[2,3], S. Satapathy[1,3*] and S. K. Majumder[1,3]

[1]*Laser Biomedical Applications Division, Raja Ramanna Centre for Advanced Technology, Indore, 452013, Madhya Pradesh, India.*

[2]*BARC Beamlines Section, Raja Ramanna Centre for Advanced Technology, Indore 452013, Madhya Pradesh, India.*

[3]*Homi Bhabha National Institute, Training School Complex, Anushakti Nagar, Mumbai 400094, Maharashtra, India.*

\*Address for Correspondence:

E-mail Address: *sonalipra8@gmail.com; srinusatapathy@gmail.com;*

# Abstract

Poly(vinylidenefluoride-co-trifluoroethylene) (P(VDF-TrFE)) based pyroelectric as well as magnetoelectric materials offer great promises for energy harvesting for flexible and wearable applications. Hence, this work focus on solar energy harvesting as well as magnetoelectric phenomenon in two phase nanocomposite film where the constituting phases are manganese ferrite ($MnFe_2O_4$) nanoparticles and P(VDF-TrFE) polymer. Composite films have been prepared using solution casting technique. X-ray diffraction result shows higher crystallinity of these films. The ferroelectric, magnetic and magnetoelectric properties in variation with applied field and volume percentage of ferrite nanoparticles have been investigated. The preparation condition was optimized in such a way that it results improved ferroelectric polarization of nanocomposite film after incorporation of small amount of ferrite nanoparticles. The maximum magnetoelectric-coupling coefficient of about 156 mV/Oe-Cm was obtained for optimum nanocomposite film when DC bias field was applied perpendicular to electric polarization direction. From a pyroelectric device perspective, solar energy harvesting is also reported. An open circuit voltage of 5V and short circuit current of order of ~1 nA is demonstrated without any pre amplification. Hence, the combination of magnetoelectric and pyroelectric properties of nanocomposite film presented here indicate as a perfect candidate for smart materials, spintronics devices and specified magnetoelectric-based applications.

# 1. Introduction

Energy harvesting is referred as energy extraction which allows the capturing of unused ambient energy such as vibration, light, temperature variations, strain and converting into usable electrical energy. Energy harvesting is a perfect match for low-power portable microelectronics which depend on a battery power. It can provide cost-effective and environmental friendly solutions for low power applications. The conversion of thermal energy has gained more attention in the recent years.

The field of multiferroics has greatly expanded in the last few years, particularly with the discovery of magnetoelectric (ME) effect [1,2]. The requisite for the observance of this effect is the coexistence of coupling between magnetic and electric order parameters. Such ME effect has mutual control on their coupled electrical polarization and magnetization. ME response in single-phase ME materials is very weak and happening only at very low temperatures. Therefore, they cannot be implemented in technological device applications [3]. Due to limitations in single-phase material, however, composite materials which mainly consist of two phases (piezoelectric and magnetostrictive phase) have focused here.

The fascinating properties of P(VDF-TrFE) such as light weight, large pyro-and piezoelectric efficiency, high elasticity, transparency and flexibility attract for potential applications [4,5]. Even if PVDF shows low dielectric constant, P(VDF-TrFE) have high dielectric constant which leads to good magnetoelectric coupling property [6–8]. Therefore, designing a suitable polymer magnetoelectric nanocomposite material is a subject of intensive research for promising technological applications. Experimental research have shown that piezoelectric polymers in comparison to ceramic-based composite ME materials have solved some problems such as fragility, low electrical resistance and high dielectric losses [9,10]. From a pyroelectric device perspective, ferroelectric P(VDF-TrFE) is a fascinating material for harvesting thermal energy. This extends the range of potential applications to the biomedical, energy, power, and signal processing fields. In this regard, the conversion of mechanical and thermal energies into electrical energy by exploiting piezoelectric and pyroelectric materials has attracted considerable attention [11,12].

Many studies on ME properties in ceramic/polymer composites, such as PZT/P(VDF-TrFE), Ni/P(VDF-TrFE) and alloy/polymer based composites have limitations to some extent when used in nanocomposite due to the large leakage current and easy oxidization [13,14]. As a replacement, oxide based magnetostrictive materials have been proposed for application in ME nanocomposites. Among different magnetic oxide materials, $MnFe_2O_4$ is an economic alternative to the existing alloy-based magnetostrictive materials. $MnFe_2O_4$ has magnetostrictive coefficients almost -55 ppm with high Curie temperature [15,16]. Ferrite nanoparticles are fascinating materials due to their chemical and thermal stability and unique structural, optical, magnetic, electrical, and dielectric properties [17]. They have wide potential technological applications in high density magnetic recording and switching devices [18–20], sensor technology, photoluminescence, biosensors, magnetic drug delivery, permanent magnets, magnetic refrigeration, magnetic liquids, microwave absorbers, biomedicine (hyperthermia) [21,22]. Bulk Manganese ferrite ($MnFe_2O_4$) has a partially inverse spinel structure with 20% of $Mn^{2+}$ ions located at octahedral (B) sites and 80% located at tetrahedral sites [23]. $MnFe_2O_4$ nanoparticles has attracted due to its good biocompatibility, high magnetization value, high anisotropy, size-dependent saturation magnetization, heat-resistant, environmentally friendly, non-toxic, high shock resistant [24].

In this context, we report fabrication of $MnFe_2O_4$/P(VDF-TrFE) 0–3 nanocomposite films with good dielectric, ferroelectric, magnetic, magnetoelectric as a product response of $MnFe_2O_4$ nanoparticles and P(VDF-TrFE) and finally pyroelectric properties at room temperature. The concentration of $MnFe_2O_4$ nanoparticles significantly influences the ferroelectric properties and hence ME and response of the copolymer matrix. The optimized nanocomposite films show high coupling coefficient, which is first reported value in P(VDF-TrFE) matrix. In this work we implemented lock in technique which was reported by G. V. Duong et al [25]. Transverse and longitudinal magnetoelectric coupling of electrically poled films are reported which were measured by keeping magnetization (M) and polarization (P) in perpendicular and parallel manner, respectively. The variation of magnitudes of α with volume % of $MnFe_2O_4$ and DC bias field is also emphasized. Additionally, solar energy harvesting property of this nanocomposite film was also tested.

## 2. Experimental

### 2.1. Materials and methods

Manganese ferrite ($MnFe_2O_4$) nanoparticles have been synthesized via sol-gel auto combustion route. The starting materials were Manganese nitrate tetrahydrate (Mn $(NO_3)_2 \cdot 4H_2O$) (purity 99.9985%, Alfa Aesar) and iron nitrate nanohydrate (Fe $(NO_3)_3 \cdot 9H_2O$) (purity 99.99%, Alfa Aesar). These metal nitrates were dissolved in distilled water with continuous stirring to get a clear solution. Then citric acid (Alfa Aesar) was added under continuous stirring and the solution was maintained at 70°C for 1 hour. Continuous heating at 200°C leads to brown color gel and finally results to formation of manganese ferrite powders. Then the powders were grounded into fine powders.

The nanocomposite films were prepared using doctor's blade method [26]. The required amount of P(VDF-TrFE) (70:30) powder (Poly K) was first dissolved in *N, N*-dimethyl formamide (DMF) (Fluka) at 70°C using the hot plate assembled with mechanical Teflon stirrer. Continuous stirring of the solution for nearly two hours leads to complete dissolution of the P(VDF-TrFE) powder. Different volume % (0.5, 1 and 2) of manganese ferrite powders were added to the solution and again continuously stirred for two hours at 100°C. The solution was ultra-sonicated for few hours to ensure the uniform dispersion of magnetic nanoparticles. After that, the solution was casted on a clean glass substrate into a film of uniform thickness using doctor's blade by keeping constant velocity and gap between glass substrate and blade. The solvent was evaporated in an oven at 100°C, resulting in a film with a thickness of 40 $\mu m$.

## 2.2 Characterization techniques

The phase of the $MnFe_2O_4$ nano powder and $MnFe_2O_4$/P(VDF-TrFE) nanocomposite films was confirmed by X-ray diffractometer (Rigaku Geigerflex) with Cu K$\alpha$1 (wavelength $\lambda$ = 1.54 Å) as the radiation source and FTIR spectrometer (JASCO-660 plus). The size of the nanoparticles and the microstructure of composite films were examined by Carl Zeiss (Sigma-02) field emission scanning electron microscope (FESEM). The X-ray absorption near-edge structure (XANES) were collected in transmission mode at the synchrotron beamline-09 RRCAT, Indore (India) (energy range 4–25 keV, energy resolution 10 keV). Spectra at the Mn (6539 eV) and Fe (7112 eV) K-edges were acquired at room temperature using a Si (111) double crystal monochromator. Novo-control Alpha impedance analyzer was used for the dielectric studies of the P(VDF-TrFE) and $MnFe_2O_4$/P(VDF-TrFE) nanocomposite films. Ferroelectric properties of these films were examined using P–E loop tracer (Marine India) and Magnetic properties of the $MnFe_2O_4$ particles and $MnFe_2O_4$/P(VDF-TrFE) nanocomposite films were studied using S700X SQUID magnetometer (Cryogenics Ltd, UK). Magnetoelectric measurement was carried out using lock in amplifier method (Marine India). Solar energy harvesting study was demonstrated with the help of solar simulator (Enli Tech.,Taiwan) as radiation source and Keithley 2450 source meter.

## 3. Results and Discussions

### 3.1. Structural analysis

The XRD spectra of the MnFe$_2$O$_4$ nanoparticles are shown in fig.1 (a). All of the main peaks are indexed as the spinel structure of MnFe$_2$O$_4$ according to JCPDS file #731964 (fig.1 (b)). Fig.1 (d) shows well-fitted Rietveld refinement of ($\chi^2$=1.87) MnFe$_2$O$_4$ nanoparticles. The peak profiles were fitted using Pseudo-Voigt function in Fig. 1(d). The result confirms a cubic spinel structure with space group symmetry Fd-3m. It is seen that all the nanoparticles show a single-phase as no other secondary phase is found in the refinement. The unit cell parameters, atomic position, bond length, fitting factor and tolerance factor obtained from refinement are summarized in Table 1. Crystal structure of MnFe$_2$O$_4$ is plotted by Visualization of Electronic Structural Analysis (VESTA) software. Figure 1(e) shows the 3D lattice structure of MnFe$_2$O$_4$, which is cubic spinel type. The blue, red and green color balls in figure denotes Mn, Fe and Oxygen ions, respectively. Tolerance factor is a useful tool to evaluate the phase stability of a certain crystal structure. The tolerance factor ($\tau$) was measured to be 0.82 by using the formula

$$\tau = \frac{\sqrt{3}}{2} * \left(\frac{R_B + R_X}{R_A + R_X}\right) \qquad (1)$$

where, $R_A$, $R_B$ and $R_X$ are ionic radius of A-site, B-site and oxygen ion, respectively. As the calculated tolerance factor value is less than one, this predicts the phase stability of synthesized spinel structure [27]. The discrepancy (less than 1) in tolerance factor is because the octahedral are compressed to constitute the main framework of spinel structure.

The average crystallite size of MnFe$_2$O$_4$ nanoparticles was calculated to be ~16 nm from X-Ray line broadening by using Scherer's equation

$$t = \frac{k\lambda}{\beta \cos\theta} \qquad (2)$$

where, t is average crystallite size, $\lambda$ is the wavelength of the Cu-K$\alpha$ radiation ($\lambda$= 1.5406Å), $\theta$ is the Bragg's diffraction angle, and $\beta$ is the full width at half maximum (FWHM) of the peak in the X-ray diffraction pattern [28]. Figure 1(c) represents the plots of XRD patterns of the MnFe$_2$O$_4$/P(VDF-TrFE) nanocomposite films having different volume fraction of MnFe$_2$O$_4$ in P(VDF-TrFE). Highly intense diffraction peak at ~20° confirms the presence of $\beta$-phase in P(VDF-TrFE) film. It is clear from the XRD patterns of MnFe$_2$O$_4$/P(VDF-TrFE) nanocomposite films that β-phase P(VDF-TrFE) exists in all nanocomposite films. However, the intensity of P(VDF-TrFE) peak dominates over the manganese ferrite peak in the composite films. This is due to the low content of manganese ferrite nanoparticles in the composite, in comparison to P(VDF-TrFE) content. The nano size of the manganese ferrite powders as well as the distribution of nanoparticles in the polymer is further confirmed using FESEM measurement. The size of the MnFe$_2$O$_4$ nanoparticles was determined with the help of ImageJ software. The data obtained from the software was fitted with a lognormal function. Finally, we found a mean particle diameter approximate to 17 nm from the fitted data (inset of Fig.2 (a)). This mean diameter is well matched with the obtained particle size from XRD. The evenly dispersion of nano particles is also observed from the FESEM image of nanocomposite. The spherulitic type microstructure with self-connected continuous network of P(VDF-TrFE) was noticed in nanocomposite.

| Structural parameters of MnFe$_2$O$_4$ nanoparticles | Values obtained from Rietveld Refinement |
|---|---|
| Space group | F d -3 m (227) |
| Lattice type | F |
| Structure | cubic |
| a = b = c (Å) | 8.3921 |
| V(Å$^3$) | 591.0502 |
| α = β = γ (degrees) | 90.00 |
| Mn (x,y,z) | (0.0010, 0.0010, 0.0010) |
| Fe (x,y,z) | (1.6355, 1.6355, 1.6355) |
| O (x,y,z) | (0.7600, 0.7600, 0.7600) |
| d$_{<Mn-O>}$ (Å) | 2.0300 |
| d$_{<Fe-O>}$ (Å) | 1.8100 |
| R$_p$ (%) | 16.9 |
| R$_{wp}$ (%) | 15.9 |
| R$_{exp}$ (%) | 11.65 |
| $\chi^2$ | 1.87 |
| τ | 0.82 |

***Table 1.*** *Crystallographic information of MnFe$_2$O$_4$ nanoparticles obtained from Rietveld Refinement at room temperature*

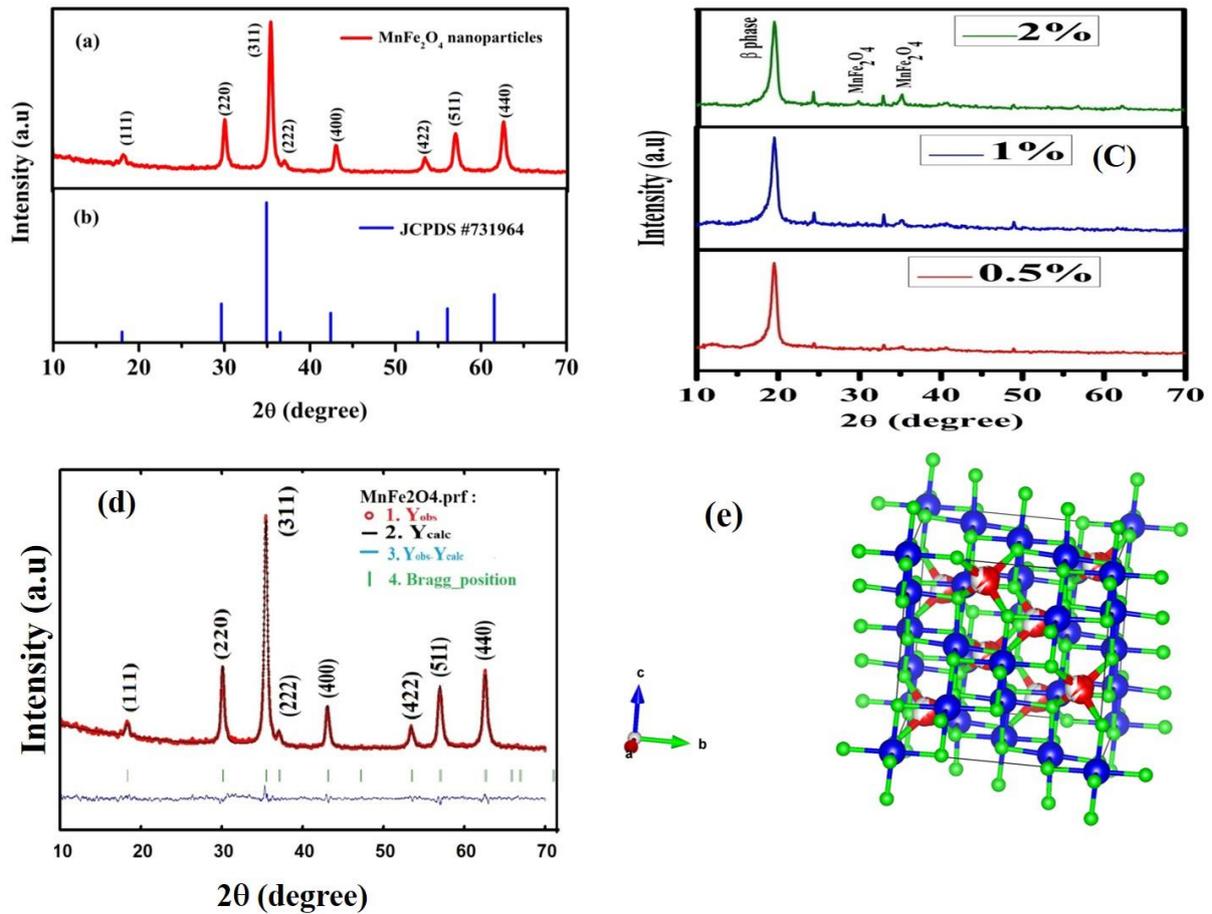

***Fig.1.*** *(a) X-ray diffraction pattern and (b) JCPDS file of MnFe$_2$O$_4$ nanoparticles; (c) X-ray diffraction pattern of 0.5, 1 and 2 volumes % MnFe$_2$O$_4$/P(VDF-TrFE) nanocomposite films; (d) Rietveld refinement profile of MnFe$_2$O$_4$ nanoparticles; (d) Crystal structure of MnFe$_2$O$_4$ generated from the Rietveld refinement of XRD data using VESTA software.*

### 3.2. FTIR studies

Formation of the spinel MnFe$_2$O$_4$ nanoparticles was further supported by the wavelength dependent transmittance data obtained using a FTIR spectrometer. Figure 2(c) shows a FTIR spectrum below 1000 cm$^{-1}$ with common features of ferrites. The peak observed at 646 cm$^{-1}$ is assigned to the intrinsic stretching vibration of the metal-oxygen at the octahedral and tetrahedral sites [15]. Peak at 1384 cm$^{-1}$ can be assigned to the carboxylate group on the MnFe$_2$O$_4$ nanoparticles due to citric acid [29]. The presence of band at 1640 cm$^{-1}$ confirms the stretching vibration of -CH$_2$ while the broad band present at 3356 cm$^{-1}$ is due to O-H bond of water vapor present in air. In summary, the spectroscopic studies confirm the formation of MnFe$_2$O$_4$ nanoparticles. In order to further confirm the presence of both Manganese ferrite and P(VDF-TrFE) in nanocomposite film, FTIR transmittance data for composite films were collected and compared with the data of pure P(VDF-TrFE) film. The peaks at 848 and 1289 cm$^{-1}$ confirm the β-phase of P(VDF-TrFE) in nanocomposite film. It is important to notice that those two peaks which is due to β-phase of P(VDF-TrFE) is less intense in pure P(VDF-TrFE) but enhanced in nanocomposite film. Hence, in order to compare the percentage fraction of β-phase, the relative fraction of β-phase was quantified as 79% and 97% for pure P(VDF-TrFE) and 0.5 vol % nanocomposite, respectively using the formula

$$F(\beta) = \frac{A(\beta)}{\frac{K(\beta)}{K(\alpha)} * A(\alpha) + A(\beta)} \quad (3)$$

where, A(β) and A(α) are the absorbance at 848 and 764 cm$^{-1}$, respectively, K(β) and K(α) are the absorption coefficients at the respective wave numbers, whose values are 7.7 × 10$^4$ and 6.1 $\times 10^4$ cm$^2$ mol$^{-1}$, respectively. This confirms that inclusion of ferrite nanoparticles into polymer matrix promotes the growth of β crystalline phase in nanocomposite. In addition to this, the carboxylate group on the MnFe$_2$O$_4$ nanoparticles which is denoted by the peak at 1342 cm$^{-1}$ is also present in nanocomposite and not in pure P(VDF-TrFE). Hence IR spectra confirms the presence of both the phases in nanocomposite films.

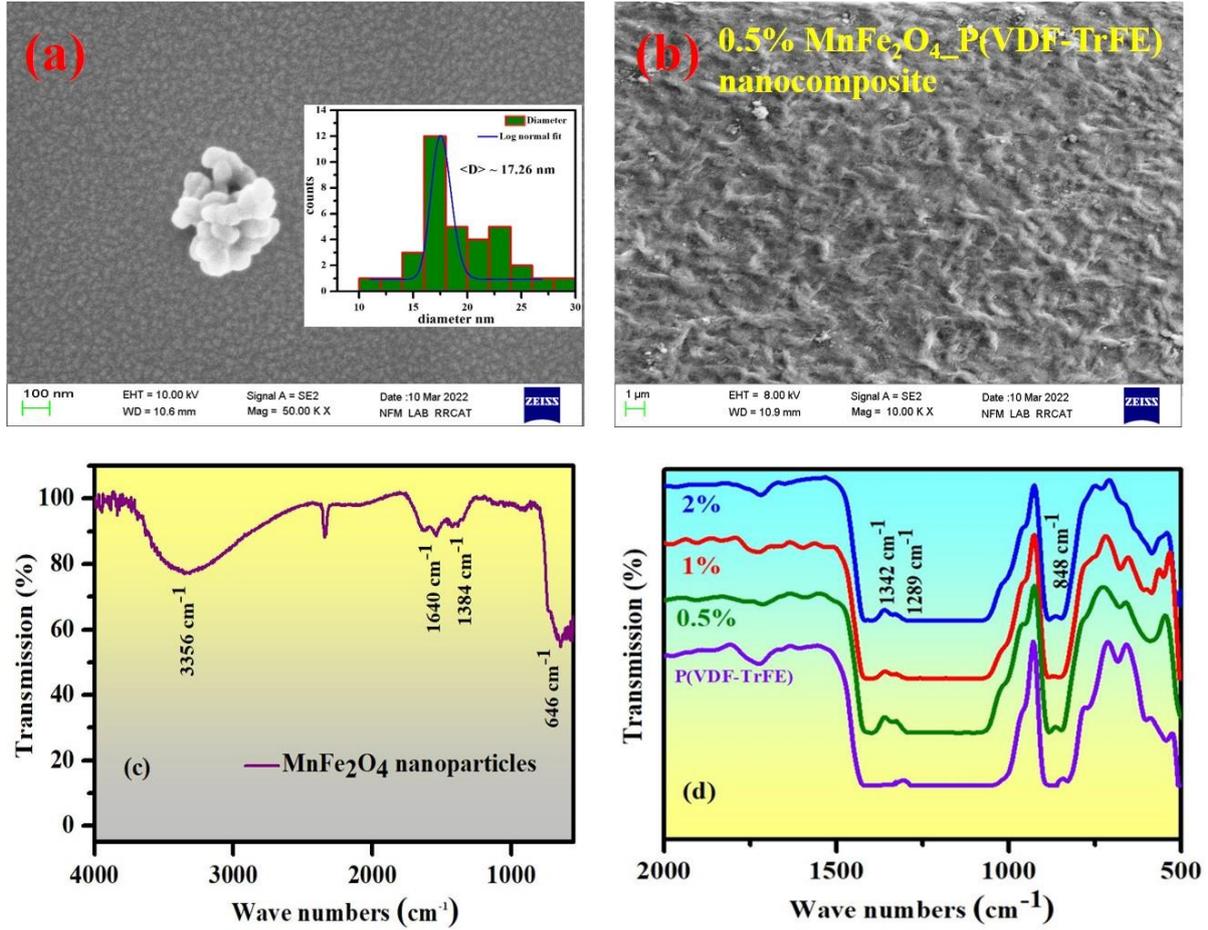

***Fig.2.*** *(a) and (b) FESEM image of $MnFe_2O_4$ nanoparticles and 0.5 vol % $MnFe_2O_4$/P(VDF-TrFE) nanocomposite film, respectively; FTIR spectrum of (c) $MnFe_2O_4$ nanoparticles and (d) pure P(VDF-TrFE) & $MnFe_2O_4$/P(VDF-TrFE) nanocomposites*

### 3.3. XANES spectroscopy

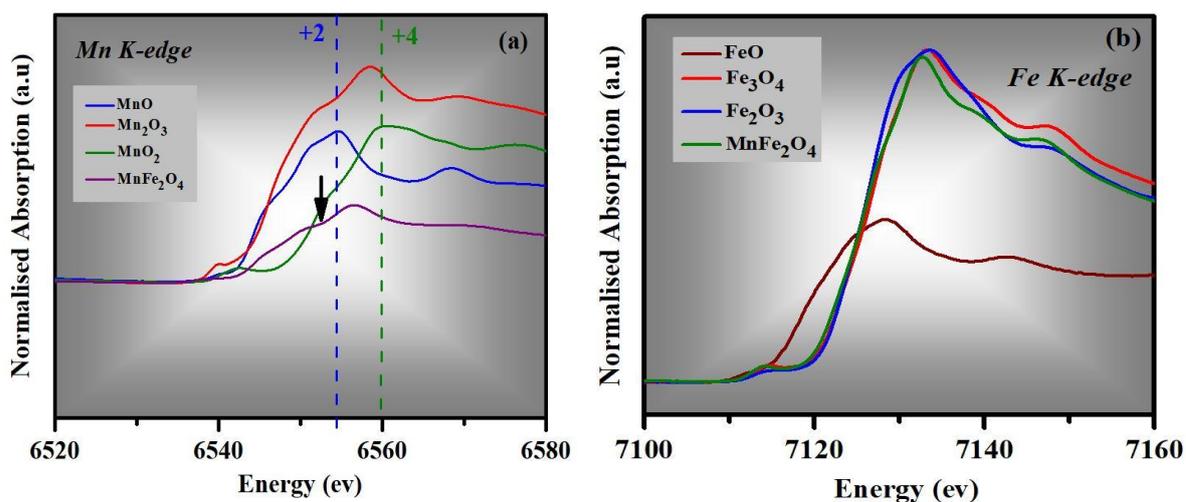

***Fig.3.*** *Normalized XANES spectra recorded at (a) Mn K-edge and (b) Fe K-edge of MnFe$_2$O$_4$ nanoparticles*

The cation distribution between tetrahedral and octahedral sites is an important point to study as it influences the magnetic properties of MnFe$_2$O$_4$ nanoparticles. The bonding environment of Mn and Fe was studied by means of X-ray absorption fine structure spectroscopy. Generally, ferrite nanoparticles prepared at lower temperatures (here combustion method) present a non-equilibrium cation distribution. Figure 3 shows the room temperature XANES spectra for the as synthesized MnFe$_2$O$_4$ nanoparticles recorded at the Mn and Fe K-edges, respectively and in the same conditions for the standard oxides. The XANES spectra were processed in the usual way to obtain normalized absorbance. Information on the oxidation state of Mn and Fe in MnFe$_2$O$_4$ nanoparticles has been obtained by comparing the main absorption edge position with those of standard reference oxides. The Mn K-edge XANES spectra (Fig. 3a) shows that the absorption edge of MnFe$_2$O$_4$ nanoparticles is in between the position of MnO and MnO$_2$. This indicates that the average oxidation state of Mn in as synthesized MnFe$_2$O$_4$ nanoparticles is higher than 2+, presenting as mixed Mn$^{2+}$ and Mn$^{3+}$ oxidation state. It is also noticed a shoulder peak (black vertical arrow) at around 6552 eV for MnFe$_2$O$_4$ nanoparticles. It is related to the Jahn−Teller effect, which results in an elastic distortion of the octahedral sites in order to energetically compensate rearrangement of the electronic distribution of octahedrally coordinated $Mn^{3+}$ cations [30,31]. In Fig. 3b, the XANES spectra of the $MnFe_2O_4$ nanoparticles at the Fe K-edge are compared with those of the standard FeO, $Fe_3O_4$ and α-$Fe_2O_3$. The absorption edge position of the $MnFe_2O_4$ nanoparticles is almost match with standard $Fe_3O_4$ while slightly less than the edge position of α-$Fe_2O_3$, indicating the presence of both $Fe^{2+}$ and $Fe^{3+}$ in synthesized $MnFe_2O_4$ nanoparticles. It was also noticed that the intensity of the pre-edge peak is very similar in $MnFe_2O_4$ nanoparticles and α-$Fe_2O_3$. The presence of pre-edge peak suggests a fraction of $Fe^{3+}$ in non-symmetrical sites (tetrahedral site) i.e., the nanoparticles have degree of inversion. This is because tetrahedral symmetry is non-centrosymmetric and it enables the characteristic pre-edge peak due to 1s→3d transitions which become dipole allowed after mixing of the metal d with 4p states [32,33].

To determine the valence state of Mn and Fe in $MnFe_2O_4$ nanoparticles, the XANES spectra of the standard oxides were analyzed by applying Linear Combination Fitting (LCF) method to the spectra of $MnFe_2O_4$ nanoparticles. In this calculation, the XANES spectrum of $MnFe_2O_4$ nanoparticles at Mn K-edge was analyzed considering a linear combination of XANES spectra obtained from MnO, $Mn_2O_3$ and $MnO_2$. Similarly, in Fe K-edge, a linear combination of XANES spectra of $Fe_3O_4$ and FeO were considered for LCF. The quality of the fit corresponds to low R factors (< 1%). The LCF fit result in Mn K-edge (Fig.4a) shows nearly 4% MnO, 78% $Mn_2O_3$ and 18% $MnO_2$ in synthesized nanoparticles. The oxidation state of Mn in these oxides is like +2, +3 and +4 respectively. Hence, this result indicates that Mn has mixed valence states in $MnFe_2O_4$ nanoparticles where $Mn^{3+}$ state is more than other $Mn^{2+}$ and $Mn^{4+}$ states. Further, in Fe K-edge, the LCF fit (Fig.4b) points out the proportions like 96% $Fe_3O_4$ (mixture of +2 and +3 oxidation states) and 4% FeO (+2 oxidation state). Such proportions lead that there is 52% $Fe^{2+}$ and 48% $Fe^{3+}$ state in $MnFe_2O_4$ nanoparticles. As the nanoparticles are synthesized at low temperature method, the thermal energy is too low to overcome the energy barrier to an ordered cation distribution. This change of cation distribution is consistent with the change of unit cell size. It was observed that the lattice constant obtained from Rietveld refinement which is about 8.39 Å is less than the bulk lattice constant 8.51 Å. It was previously studied that the inverse spinel has a smaller unit cell than its counterpart with normal cation occupancy [34,35]. This difference in unit cell size has been attributed to a shorter cation-anion bond for $Fe^{2+}$ in the tetrahedral sites compared to $Fe^{3+}$ in the octahedral sites, and to the possible change of Mn

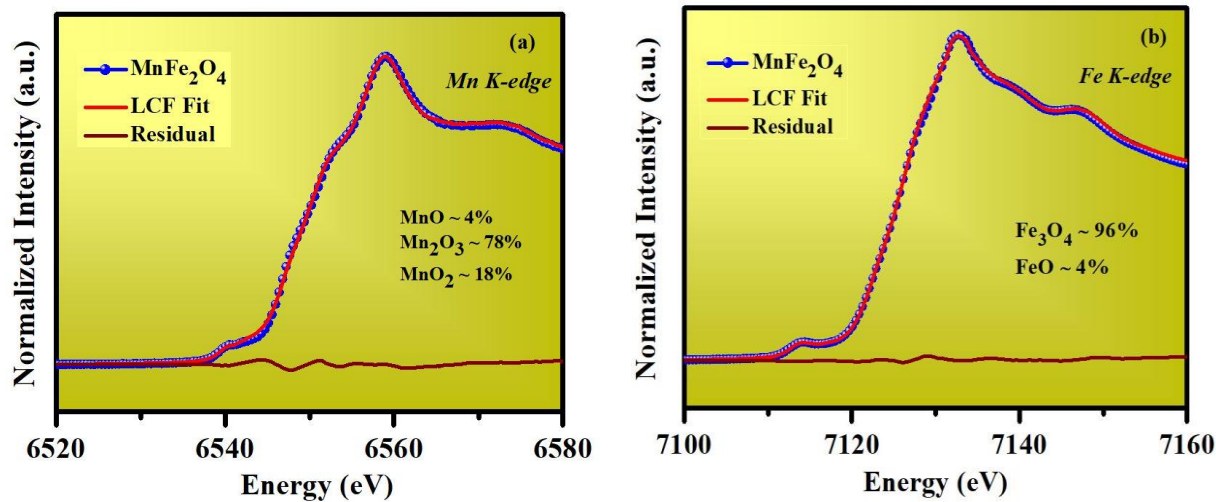

***Fig.4.*** *Experimental XANES spectra at (a) Mn K-edge and (b) Fe K-edge of MnFe$_2$O$_4$ nanoparticles (blue open spheres) fitted (red full line) using LCF method. Residual is also shown (wine full line)*

oxidation state. Hence according to the result, the synthesized MnFe$_2$O$_4$ nanoparticles is in mixed state with partially inverted.

## 3.4. Dielectric studies

Figure 5(a) represents the real part of dielectric constant ($\varepsilon^{/}$) of pure P(VDF-TrFE) and MnFe$_2$O$_4$/P(VDF-TrFE) nanocomposites in the frequency range of 100 Hz to 1 MHz at room temperature. The permittivity of nanocomposite films enhances from pure P(VDF-TrFE) after incorporation of manganese ferrite nanoparticles. This is attributed to increase in interfacial polarization between polymer matrix and manganese ferrite nanoparticles. However, the permittivity was found to be independent of frequency in higher frequency region. The frequency variation of permittivity at different temperatures (in the range from 30°C to 100°C with 10°C interval) for 0.5%, 1% and 2% MnFe$_2$O$_4$/P(VDF-TrFE) nanocomposite films are shown in fig. 5(b), (c) and (d), respectively. It is realized that permittivity of nanocomposite films increases with increase in temperature due to easy orientation of dipoles as the polymer chains breaks with temperature. This trend is also observed at a particular frequency as shown in figures 6(a), (b) and (c). In low frequency region (figures 5(b), (c) and (d)), the increase in permittivity with increase in temperature becomes notable. This is mainly attributed to Maxwell-Wagner-Siller (MWS) interfacial polarization. Likewise, as the temperature increases, it expands the interface of polymer and nanoparticles, which ultimately increases the interfacial polarization.

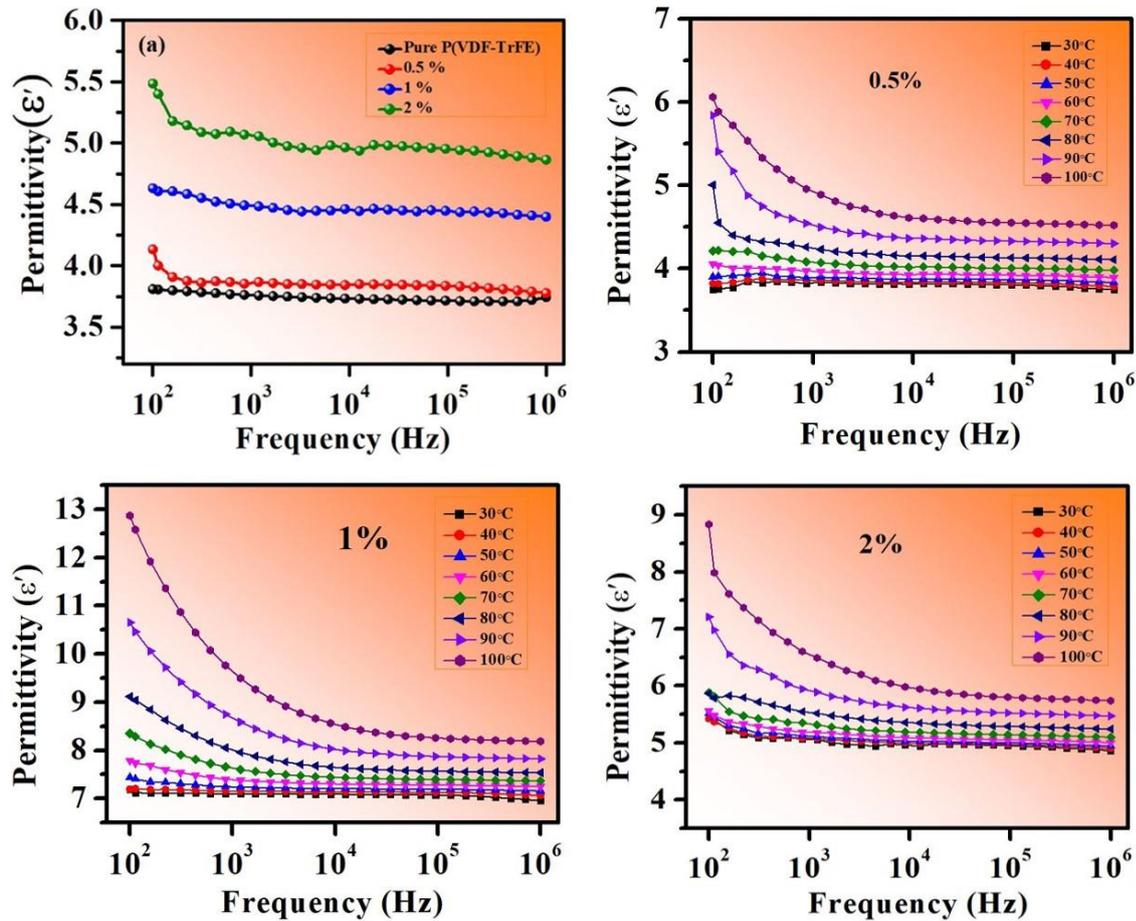

***Fig.5.*** *(a) Room temperature frequency dependent permittivity of pure P(VDF-TrFE) and MnFe$_2$O$_4$/P(VDF-TrFE) nanocomposite films; (b), (c) and (d) frequency dependent permittivity of 0.5, 1 and 2 volume % nanocomposite film from RT to 100°C, respectively.*

### 3.5. Ferroelectric Properties

The room temperature spontaneous polarization of pure P(VDF-TrFE) and MnFe$_2$O$_4$ /P(VDF-TrFE) nanocomposites were recorded at frequency 50 Hz (fig. 6(d)). A well-defined saturated loop was observed for pure P(VDF-TrFE) at application of 483 kV cm$^{-1}$ electric field resulting to saturation polarization nearly 5.6 μC cm$^{-2}$ at coercivity 310 kV cm$^{-1}$. The observed ferroelectric hysteresis loop for 0.5 vol % nanocomposite indicates enhanced ferroelectric properties. The ferroelectric properties retain up to 0.5 vol % of manganese ferrite nanoparticles in P(VDF-TrFE) polymer. The saturation polarization was found to be increased from pure P(VDF-TrFE) to the value 8.2 μC cm$^{-2}$ with E$_C$ = 400 kV cm$^{-1}$ at application of electric field555 kV cm$^{-1}$. It is concluded that uniform distribution of nanoparticles in polymer matrix assists well-saturated ferroelectric loops of 0.5 vol % composite, which is also consistent with nanoparticles. Hence, enhanced saturation polarization and coercive field in compare to pure P(VDF-TrFE) were observed in 0.5 vol % nanocomposite. As the concentration of nanoparticles increases in polymer, we observed decrease in saturation polarization. This may be due to hindrance of nanoparticles in polymer chain structure, resulting to increase in conductivity in composite.

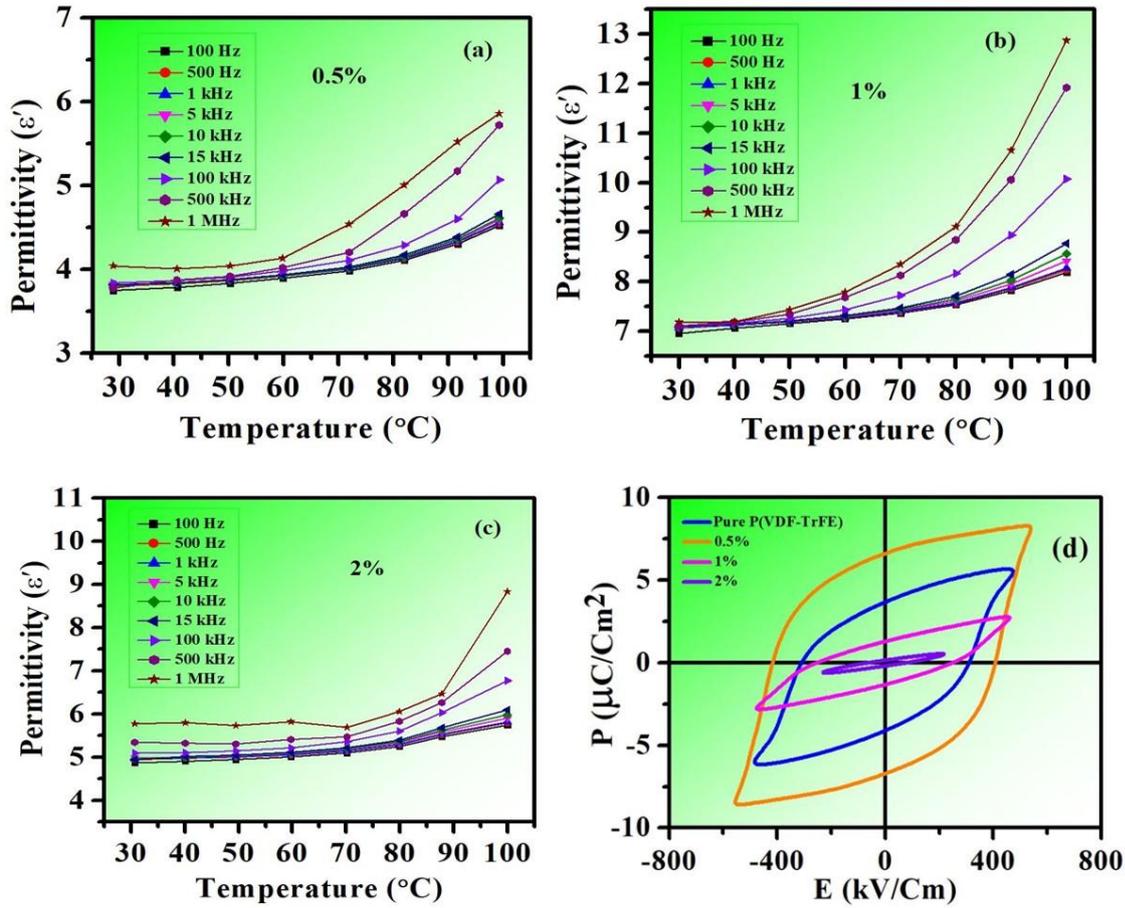

***Fig. 6.*** *(a), (b) and (c) Temperature dependent permittivity of 0.5, 1 and 2 volume % nanocomposite film in the frequency range 100 Hz to 1MHz, respectively; (d) P-E hysteresis loop measurement of pure P(VDF-TrFE) and MnFe$_2$O$_4$/P(VDF-TrFE) nanocomposite films*

### 3.6. Magnetic characterizations

To explore the impact of particle morphologies and phase compositions on the magnetic properties, MnFe$_2$O$_4$ nanoparticles were investigated with S700X SQUID magnetometer. The DC magnetization plot measured from 3 to 370 K is shown in Fig. 7(a). The data were collected (at 300 Oe) in zero field cooled (ZFC) and field cooled (FC) protocol. Fig. 7(b) exhibits the room temperature M-H hysteresis of MnFe$_2$O$_4$ nanoparticles over a magnetic field range of

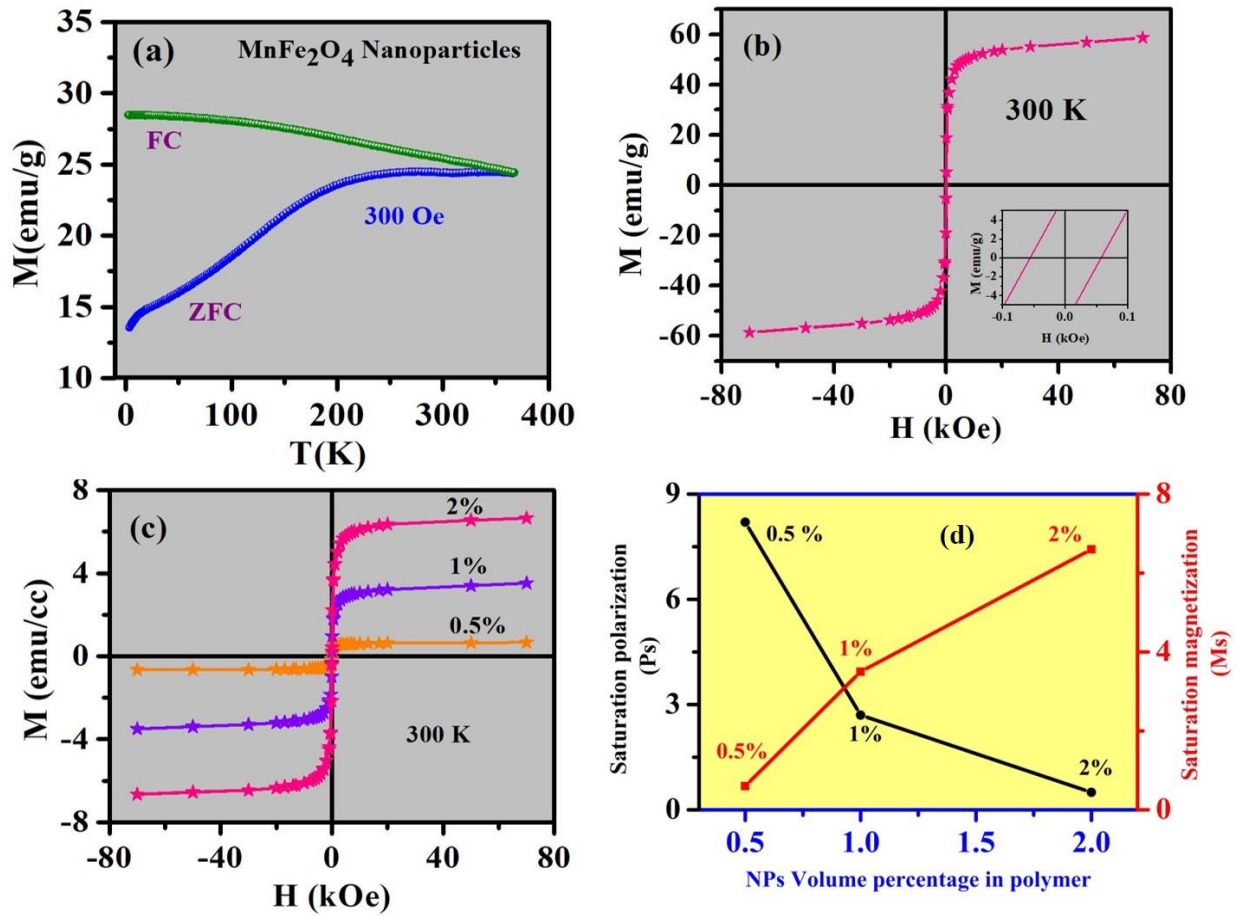

***Fig.7.*** *(a) Temperature and (b) magnetic field dependent magnetization curve of $MnFe_2O_4$ nanoparticles; (c) M-H hysteresis loop of $MnFe_2O_4$/P(VDF-TrFE) nanocomposite films; (d) comparison between saturation polarization and saturation magnetization of nanocomposite films with concentration of nanoparticles in P(VDF-TrFE)*

±70 kOe. It is evident from figure that $MnFe_2O_4$ nanoparticles show ferrimagnetic behavior with saturation magnetization nearly 58 emu/g (~ 2.2 $\mu_B$ per formula unit) i.e., nearly about 73% inverse spinel and low coercive field i.e., nearly 55 Oe. This high percentage of inverse spinel is due to high percentage of $Fe^{3+}$ ions in tetrahedral site and partial oxidation of Mn ions from +2 to +4 states confirmed from XANES analysis. Typically, the inverse parameter in ferrite nanoparticles varies between 0.2 to 0.9 which is highly depend on different synthesis method, particle size, preparation condition etc. which ultimately causes different cation distribution in tetrahedral and octahedral sites. Fig. 7(c) shows the room temperature M-H of $MnFe_2O_4$/P(VDF-TrFE) nanocomposites. From figure, it is clear that the composites reserve the magnetic properties of $MnFe_2O_4$ nanoparticles, but the saturation magnetization of the composites increases with the increase in nanoparticles concentration. This is due to the contribution of the magnetic moment of individual magnetic particle in polymer. However, the decrease of saturation magnetization of nanocomposites in comparison to $MnFe_2O_4$ nanoparticles might be because of magnetic moment of very small amount of nano particles involved per cubic centimeter and due to quenching of the surface moment of magnetic nanoparticles due to presence of non-magnetic polymer phase [36].

### 3.7. Magnetoelectric properties

To reveal the interaction among electric and magnetic dipoles, magneto-electric coupling coefficient ($\alpha_{ME}$) of the nanocomposite films was examined. In this $MnFe_2O_4$/P(VDF-TrFE) nanocomposite films, the ME effect was generated as product property between magnetostrictive and piezoelectric phases. DC magnetic field dependence of $\alpha_{ME}$ was measured by superimposing an AC magnetic field (25 Oe) at frequency 10 kHz for all nanocomposites. For the applied static magnetic field, strain is generated in magnetic phase due to magnetostrictive property of manganese ferrite. Hence, the corresponding stress is experienced by P(VDF-TrFE) through the interface. Due to the piezoelectric property of P(VDF-TrFE) polymer a voltage is induced in the composite. This is known as linear magnetoelectric effect, which is quantified by magnetoelectric voltage coefficient according to the formula

$$\alpha^V{}_{ME} = \frac{\delta V}{t * \delta H} \qquad (4)$$

where, $\delta V$ is the output voltage, t is the thickness of nanocomposite film and $\delta H$ is the AC magnetic field applied. The magnetoelectric coupling mainly depends on the ferroelectric properties of the piezoelectric phase in composite. In ferrite/polymer composite, ferroelectric polarization enhances up to a certain ferrite content depending on the electrostatic interaction between nanoparticles and polymer (also evidenced from ferroelectric loop in fig 6d and 7d). From fig 9(d) it was evidenced that the ME voltage coefficient is found to be increases up to 0.5 vol % of ferrite nanoparticles. For further high concentration of nanoparticles, the piezoelectric properties decrease because of polymer chain disruption due to nanoparticles which ultimately leads to decrease in ME voltage output. In the present study, the ME voltage coefficient was measured in two modes i.e., transverse $\alpha_V{}^{31}$ (direction between applied DC magnetic field and output voltage is perpendicular) and longitudinal mode $\alpha_V{}^{33}$ (direction between applied DC magnetic field and output voltage is parallel). The ME coefficients of 0.5%, 1%, and 2% $MnFe_2O_4$/P(VDF-TrFE) electrically poled (poled across the thickness) films are shown in the Fig. 9(a), (b) and (c) respectively. The maximum magnitude of $\alpha_V{}^{31}$ is found to be 156.14 mVOe$^{-1}$cm$^{-1}$ at 1.21 kOe DC magnetic field for 0.5% $MnFe_2O_4$/P(VDF-TrFE) poled composite film. To understand the phenomenon let's consider the plane of the film be in xy plane. Before the measurement the sample was ferroelectrically poled across the thickness or along z direction. Hence, the polarization (P) of domains is aligned along z-direction. When magnetic field is applied along x-direction, it creates maximum stress along z-direction. It is observed that in both transverse and longitudinal case, path followed by α is not identical for increasing and decreasing applied DC magnetic cycle. This is due to the fact that in the decreasing cycle of $H_{dc}$, only poled ferromagnetic domains contributes to remnant magnetization and magnetostriction, which in turn decrease the α value.

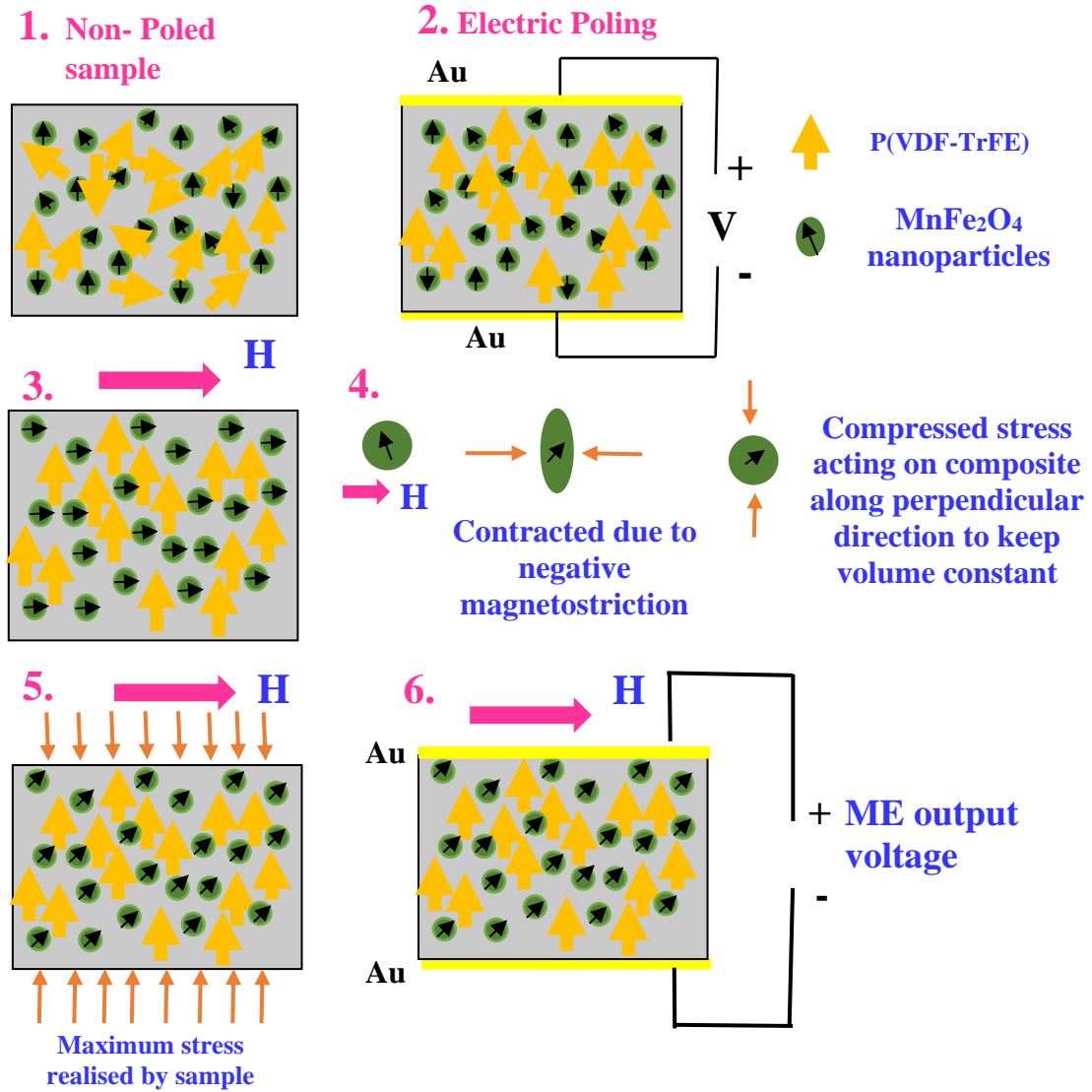

*Fig. 8. Schematic illustration of mechanism of ME coupling*

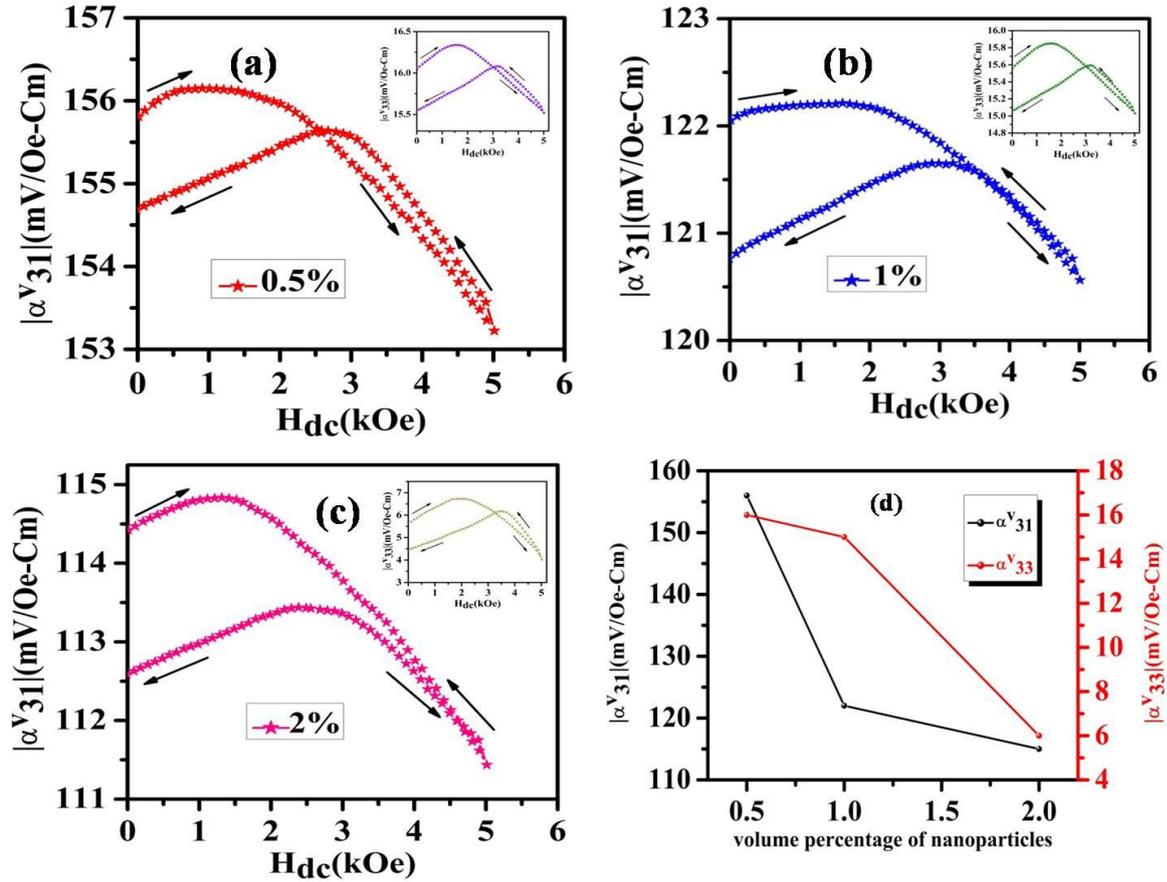

***Fig.9.*** *(a), (b) and (c) magnetoelectric voltage coefficient of 0.5, 1 and 2 volume % nanocomposite film in transverse mode (respective inset shows in longitudinal mode), respectively; (d) volume concentration dependent transverse and longitudinal magnetoelectric voltage coefficient*

It is important to notice that initially α increases with DC magnetic field and reaches a maximum and then falls with further increase of DC bias field. This is due to the fact that the output voltage is proportional to piezomagnetic coefficient (q) according to the equation

$$\alpha = q * d \qquad (5)$$

where, q is the piezomagnetic coefficient of magnetic phase and d is the piezoelectric coefficient of electric phase. In general, piezomagnetic coefficient increases with increase in the bias field $H_{dc}$. At optimum DC bias field, it attains maximum and then saturates with further increase in $H_{dc}$. Hence, α follows the trend accordingly. The behaviour of $\alpha v^{31}$ is same to that of $\alpha v^{33}$, except in magnitude ($\alpha v^{33} \sim 16$ mVOe$^{-1}$cm$^{-1}$). Since, manganese ferrite has negative magnetostriction coefficient (~ 55 ppm), nanoparticles contract along the field direction when external static magnetic field was applied.

| Sl. No | 0-3 nanocomposite | Remnant polarization ($\mu Ccm^{-2}$) | ME coefficient ($mVcm^{-1}Oe^{-1}$) | References |
|---|---|---|---|---|
| 1 | $MnFe_2O_4$_P(VDF-TrFE) | 6.5 | 156.14 | In this work |
| 2 | $BaTiO_3$_P(VDF –TrFE) | 6.03 | 22.2 | [37] |
| 3 | $CoFe_2O_4$_P(VDF-TrFE) | 17 | 42 | [38] |
| 4 | $NiFe_2O_4$_P(VDF-TrFE) | 5.8 | 90 | [39] |
| 5 | $BaFe_{12}O_{19}$_P(VDF-TrFE) | - | 15 | [40] |
| 6 | $SmFeO_3$_P(VDF-TrFE) | - | 55 | [41] |
| 7 | $NiFe_2O_4$_P(VDF-TrFE) | 1.6 | 136.4 | [42] |
| 8 | $CoFe_2O_4$_P(VDF-TrFE) | 5 | 47.1 | [43] |

*Table 2. Comparison of present work with the ferroelectric polarization and ME coefficient values of the reported 0-3 ferrite/P(VDF-TrFE) magnetoelectric nanocomposites*

To keep the volume conserved, particles have to elongate along the perpendicular direction of applied field. Therefore, maximum strain was developed in perpendicular direction of applied magnetic field and hence the ME voltage coefficient was obtained maximum for transverse mode in compare to longitudinal mode. Comparison of the ferroelectric and magnetoelectric response of the present nanocomposite with the reported 0-3 ferrite/P(VDF-TrFE) magnetoelectric nanocomposites is shown in Table 2. It is clear from the results that the present $MnFe_2O_4$/P(VDF-TrFE) nanocomposite films possess excellent magnetoelectric and ferroelectric response at room temperature, when compared with the reported other ferrite/P(VDF-TrFE) nanocomposites.

### 3.8. Solar energy harvesting

P(VDF-TrFE) is a pyroelectric material, a subclass of piezoelectric materials. This material responds to a time-varying temperature gradient and produce a change of polarization. P(VDF-TrFE) possesses spontaneous polarization due to non-centrosymmetric crystal structure. This gives rise to a permanent dipole moment along the crystal axis [44]. The working principle of pyroelectric material is; when a time varying thermal flux applied let say increasing the temperarure i.e. $dT/dt > 0$, the individual dipole will randomly align due to thermal agitation as the temperature rises up to thermal equilibrium. As a Consequence, the net polarization of the material decreases. This decrease in polarization results in the decrease in surface bound free charges which generates a current due to an electrical imbalance across the polar axis if the surfaces are connected to electric circuit. This is known as pyroelectric current. Similarly, upon reducing the temperature (dT/dt < 0), the dipoles are restored to their initial positions. This results an increase in the net polarization which leads to reverse the flow of current [45]. However, if dT/dt = 0, then no current will be generated. This builds the basis of pyroelectric energy harvesting.

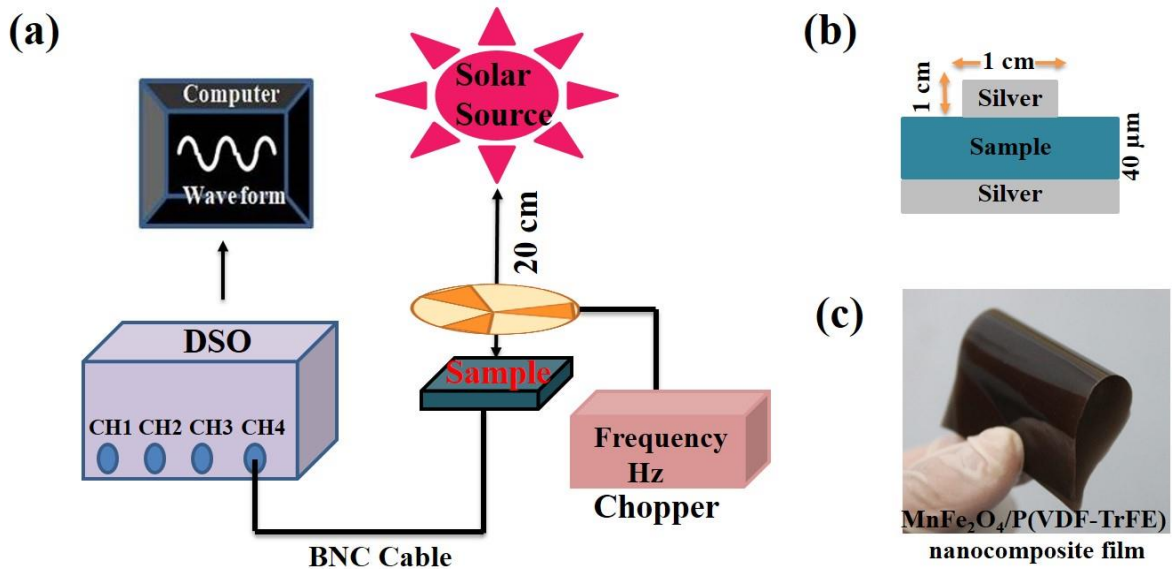

*Fig.10. (a) Schematic of pyroelectric solar energy harvesting; (b) Lateral schematic view of sample used for solar energy harvesting; (c) Picture of synthesized MnFe$_2$O$_4$/P(VDF-TrFE) nanocomposite film*

To analyse the solar energy harvesting performance of MnFe$_2$O$_4$/P(VDF-TrFE) nanocomposite film, the optimum nanocomposite film (0.5 vol % MnFe$_2$O$_4$/P(VDF-TrFE)) was selected which shows excellent ferroelectric and ME properties. Sample having surface area of 2×2 cm$^2$ with an active area of 1×1 cm$^2$ is placed normal to the incident solar radiation of photovoltaic device. Solar radiation is blocked using a chopper in order to attain cooling. During this process, cooling process happens through natural convection created by the temperature gradient. First of all, the bottom and top surface of the film were silver coated by sputtering to make a conductive medium. Thenceforth, the silver coated film electrically poled along perpendicular direction of the film. After that the film was kept at 20 cm distance normal to the solar radiation of power about 40 W/m$^2$ (using solar simulator). Connections were taken from the top and bottom surface of the sample and connected to digital storage oscilloscope (DSO) using BNC connector to measure the output voltage difference between the top and bottom electrodes.

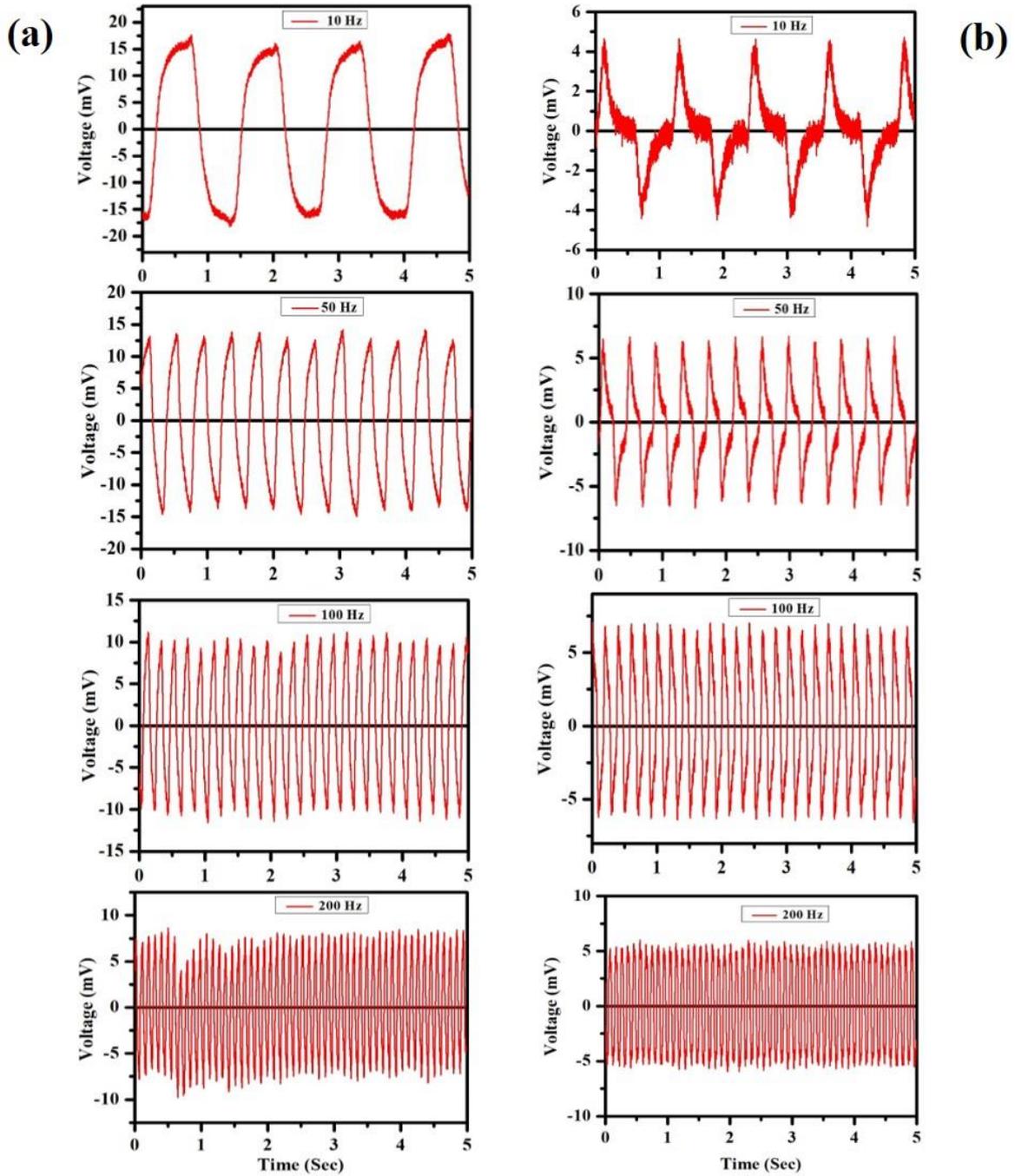

***Fig.11.*** *Pyroelectric output voltage in (a) DC and (b) AC coupled with DSO*

Fig 10(a) shows the schematic of the above arrangement. The output peak to peak was recorded across load resistance of 1 MΩ by applying different frequencies using the chopper. Figure 11 shows the generated output voltage in response to temperature change caused by chopper at frequencies 10, 50, 100 and 200 Hz. The output voltages are recorded at same time scale with DC and AC coupling mode in DSO.

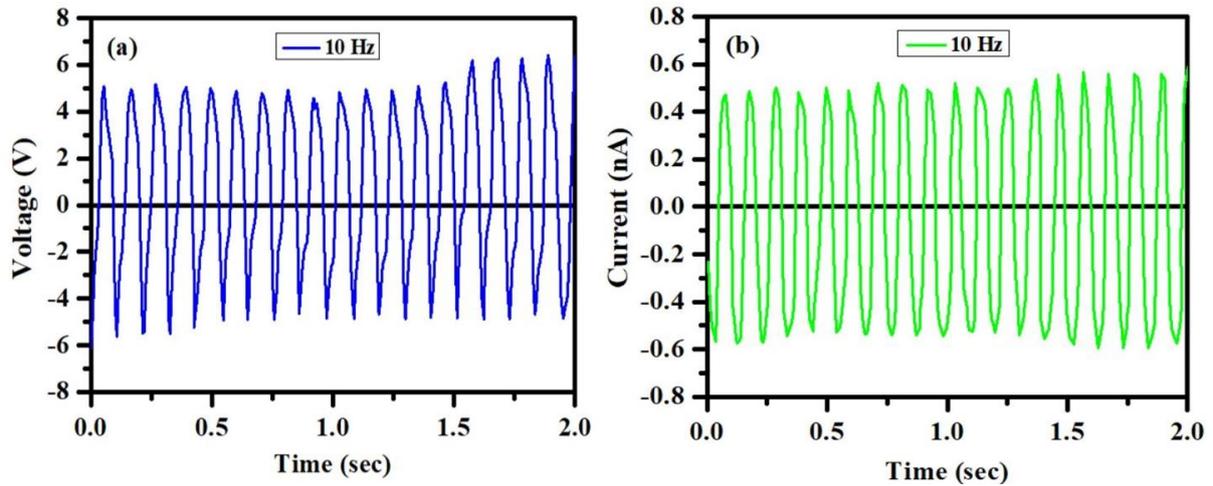

*Fig.12. Solar energy harvesting (a) open circuit voltage and (b) short circuit current at 10 Hz*

In DC coupling mode (fig. 11a), the total charge develops between two electrodes are collected as output. Hence in DC coupling, the sample behaves like a capacitor and the magnitude of voltage increases with decreasing the frequency due to increase in temperature gradient on the sample surface. Whereas, in AC coupling mode (fig. 11b), the sample behaves like current source and the output voltage follow the shape of pyroelectric current. Therefore, the output was collected across the load resistance as I×R. Further, the open circuit voltage and the short circuit current was measured using Keithley 2450 source meter **without any preamplification of current** (Figure 12). Hence, the maximum output power was measured to be 2.5 nW (equivalent to 2.5 mW for 200 TΩ output resistor). Reports are available on PVDF based pyroelectric energy harvesting with different load resistances [46–48]. *Zabek (2017) et al* demonstrated PVDF/graphene ink structure for maximum 20 V open circuit voltage at high load resistance 200 TΩ (Using Keithley high impedance analyzer). In our case, the maximum open circuit voltage of 5 V was found at load resistance 10 GΩ. This amplitude of output voltage at low load resistance (in comparison to 200 TΩ) in this magnetoelectric $MnFe_2O_4$/P(VDF-TrFE) nanocomposite film will be better than previous PVDF based reported results.

## Conclusions

A novel $MnFe_2O_4$/P(VDF-TrFE) nanocomposite film was prepared to demonstrate magnetoelectric and pyroelectric properties. $MnFe_2O_4$/P(VDF-TrFE) nanocomposite film with optimum 0.5 volume % exhibited improved ferroelectric properties when compared to pure P(VDF-TrFE) film. It is concluded that multiferroic properties of nanocomposite film depends on various factors like processing condition, microstructure, size of ferrite nanoparticles, concentration of ferrite in polymer, interface and ferroelectric properties in order to get perfect magnetoelectric coupling in ferrite/polymer nanocomposite. By optimizing all these characteristics, a transverse ME voltage coefficient of 156.4 mVcm$^{-1}$Oe$^{-1}$ at an optimum DC bias of 900 Oe is achieved which is very much encouraging for ME device applications. Further, due to the excellent ferroelectric properties, it provides rapid temperature gradient during heating, hence transforming solar energy to electrical energy through pyroelectric effect of these films.

## Acknowledgement

Authors thankfully acknowledge RRCAT (Indore) and HBNI, Mumbai (Sanction No. DAE/LBAD/5401-00-206-83-00-52, LT830006) for financial support. The authors are thankful to Shri Prem Kumar for XRD measurement and Mrs. Rashmi Singh for FESEM. Dr. Aasiya Shaikh is acknowledged for her help in FTIR measurement. One of the authors, is greatly thankful to Dr. Shilpa Tripathi, Md. Akhlak Alam and Mrs. Babita Vinayak Salaskar for XANES data analysis.

## References

[1] N Pereira, AC Lima, S Lanceros-Mendez, P Martins Materials (Basel) 13 (2020) 1-25. https://doi.org/10.3390/ma13184033.

[2] NA Spaldin, R Ramesh Nat Mater 18 (2019) 203–12. https://doi.org/10.1038/s41563-018-0275-2.

[3] A Lasheras, J Gutiérrez, S Reis, D Sousa, M Silva, P Martins, et al Smart Mater Struct 24 (2015) 1-6. https://doi.org/10.1088/0964-1726/24/6/065024.

[4] V Bharti, QM Zhang Phys Rev B - Condens Matter Mater Phys 63 (2001) 1–6. https://doi.org/10.1103/PhysRevB.63.184103.

[5] T Furukawa Adv Colloid Interface Sci 71–72 (1997)183–208.

https://doi.org/10.1016/s0001-8686(97)00017-1.

[6] K Yu, Y Niu, Y Bai, Y Zhou, H Wang Appl Phys Lett 102 (2013) 1-5. https://doi.org/10.1063/1.4795017.

[7] T Zhou, JW Zha, RY Cui, BH Fan, JK Yuan, ZM Dang ACS Appl Mater Interfaces 3 (2011) 2184–8. https://doi.org/10.1021/am200492q.

[8] QM Zhang, H Li, M Poh, F Xia, ZY Cheng, H Xu, et al Nature 419 (2002) 284–7. https://doi.org/10.1038/nature01021.

[9] P Martins, S Lanceros-Méndez Adv Funct Mater 23 (2013) 3371–85. https://doi.org/10.1002/adfm.201202780.

[10] RP Ummer, R Raneesh, C Thevenot, D Rouxel, S Thomas, N Kalarikkal RSC Adv 6 (2016) 28069–80. https://doi.org/10.1039/c5ra24602d.

[11] Q Zhang, A Agbossou, Z Feng, M Cosnier Sensors Actuators A Phys 168 (2011) 335–42. https://doi.org/10.1016/j.sna.2011.04.045.

[12] G Sebald, E Lefeuvre, D Guyomar IEEE transactions on ultrasonics, ferroelectrics, and frequency control 55 (2008) 538-551. https://doi.org/10.1109/TUFFC.2008.680.

[13] HLW Chan, PKL Ng, CL Choy Appl Phys Lett 74 (1999) 3029–31. https://doi.org/10.1063/1.124054.

[14] JX Zhang, JY Dai, LC So, CL Sun, CY Lo, SW Or, et al J Appl Phys 105 (2009) 0–3. https://doi.org/10.1063/1.3078111.

[15] K Zipare, J Dhumal, S Bandgar, V Mathe, G Shahane J Nanosci Nanoeng 1 (2015) 178–82.

[16] T Shanmugavel, SG Raj, GR Kumar, G Rajarajan Phys Procedia 54 (2014) 159–63. https://doi.org/10.1016/j.phpro.2014.10.053.

[17] P Chand, S Vaish, P Kumar Phys B Condens Matter 524 (2017) 53–63. https://doi.org/10.1016/j.physb.2017.08.060.

[18] AT Raghavender, NH Hong J Magn Magn Mater 323 (2011) 2145–7. https://doi.org/10.1016/j.jmmm.2011.03.018.

[19] ZŽ Lazarević, Č Jovalekić, A Recnik, VN Ivanovski, M Mitrić, MJ Romčević, et al J

Alloys Compd 509 (2011) 9977–85. https://doi.org/10.1016/j.jallcom.2011.08.004.

[20]  D Chen, HY Liu, L L Mater Chem Phys 134 (2012) 921–4. https://doi.org/10.1016/j.matchemphys.2012.03.091.

[21]  AB Naik, PP Naik, SS Hasolkar, D Naik Ceram Int 46 (2020) 21046–55. https://doi.org/10.1016/j.ceramint.2020.05.177.

[22]  K Asghar, M Qasim, D Das Mater Today Proc 26 (2018) 87–93. https://doi.org/10.1016/j.matpr.2019.05.380.

[23]  T Shanmugavel, SG Raj, GR Kumar, G Rajarajan Phys Procedia 54 (2014) 159–63. https://doi.org/10.1016/j.phpro.2014.10.053.

[24]  A Sivakumar, S Sahaya Jude Dhas, SA Martin Britto Dhas Solid State Sci 107 (2020) 106340. https://doi.org/10.1016/j.solidstatesciences.2020.106340.

[25]  V Loyau, V Morin, G Chaplier, M Lobue, F Mazaleyrat J Appl Phys 117 (2015) 1-12. https://doi.org/10.1063/1.4919722.

[26]  C Ribeiro, CM Costa, DM Correia, J Nunes-Pereira, J Oliveira, P Martins, et al Nat Protoc 13 (2018) 681–704. https://doi.org/10.1038/nprot.2017.157.

[27]  Z Song, Q Liu Crystal Growth & Design 20 (2020) 2014-18. https://doi.org/10.1021/acs.cgd.9b01673.

[28]  I V Minin, O V Minin Millimeter and submillimeter waves and applications: international conference (1994).

[29]  T Ahmad  Journal of the Korean Physical Society 62 (2013) 1696-1701. https://doi.org/10.3938/NPSM.63.201.

[30]  MA Denecke, W Gunßer, G Buxbaum, P Kuske Mater Res Bull 27 (1992) 507–14. https://doi.org/10.1016/0025-5408(92)90029-Y.

[31]  FH Martins, FG Silva, FLO Paula, De AG Juliano, R Aquino, J Mestnik-Filho, et al J Phys Chem C 121 (2017) 8982–91. https://doi.org/10.1021/acs.jpcc.6b09274.

[32]  G Palomino Turnes, S Bordiga, A Zecchina, GL Marra, C Lamberti J Phys Chem B 104 (2000) 8641–51. https://doi.org/10.1021/jp000584r.

[33]  M Wilke, F Farges, PE Petit, GE Brown, F Martin Am Mineral 86 (2001) 714–30.

https://doi.org/10.2138/am-2001-5-612.

[34] JB Goodenough, AL Loeb Phys Rev 98 (1955) 391–408. https://doi.org/10.1103/PhysRev.98.391.

[35] EJ Verwey, PW Haayman, FC Romeijn J Chem Phys 15 (1947) 181–7. https://doi.org/10.1063/1.1746466.

[36] P Martins, A Lasheras, J Gutierrez, JM Barandiaran, I Orue, S Lanceros-Mendez J Phys D Appl Phys 44 (2011) 1-7. https://doi.org/10.1088/0022-3727/44/49/495303.

[37] A Mayeen, MS Kala, MS Jayalakshmy, S Thomas, D Rouxel, J Philip, et al Dalt Trans 47 (2018) 2039–51. https://doi.org/10.1039/c7dt03389c.

[38] P Martins, Y V Ko. Kolen, J Rivas , S Lanceros-Mende ACS Appl Mater Interfaces 7 (2015) 15017–22. https://doi.org/10.1021/acsami.5b04102.

[39] A Mayeen, MS Kala, MS Jayalakshmy, S Thomas, J Philip, D Rouxel, et al Dalt Trans 48 (2019) 16961–73. https://doi.org/10.1039/c9dt02856k.

[40] J. Guttierrez; A. Lasheras; J. M. Barandiarán; R. Gonçalves; P. Martins; S. Lanceros-Mendez IEEE International Magnetics Conference 19 (2015). 10.1109/INTMAG.2015.7157266

[41] A Ahlawat, AA Khan, P Deshmukh, M Tripathi, MM Shirolkar, S Satapathy, et al J Mater Sci Mater Electron 30 (2019) 17765–72. https://doi.org/10.1007/s10854-019-02127-w.

[42] S Pradhan, P Deshmukh, AA Khan, A Ahlawat, SK Rai, S Satapathy Smart Mater Struct 30 (2021) 75034. https://doi.org/10.1088/1361-665X/ac0676.

[43] X Mu, H Zhang, C Zhang, S Yang, J Xu, Y Huang, et al J Mater Sci 56 (2021) 9728–40. https://doi.org/10.1007/s10853-021-05937-8.

[44] YJ Ko, BK Yun, JH Jung. Journal of the Korean Physical Society 66 (2015) 713–6. https://doi.org/10.3938/jkps.66.713.

[45] C Hsiao, S Liu Sensors 14 (2014) 22180–98. https://doi.org/10.3390/s141222180.

[46] SK Ghosh, M Xie, CR Bowen, PR Davies, DJ Morgan, D Manda Sci Rep 7 (2017) 1–13. https://doi.org/10.1038/s41598-017-16822-3.

[47] AK Batra, A Bandyopadhyay, AK Chilvery, M Thomas 5 (2013) 1–7. https://doi.org/10.3968/j.est.1923847920130502.2407.

[48] A Thakre, A Kumar, H C Song, D Y Jeong, J Ryu Sensors 19 (2019) 1-25. https://doi.org/10.3390/s19092170